\documentclass{Template}
\usepackage{graphicx}
\usepackage{dcolumn}
\usepackage{bm}
\usepackage{caption}
\usepackage{subcaption}
\usepackage{gensymb}
\usepackage{xcolor}
\usepackage{amsmath}
\usepackage{textcomp}
\usepackage{ragged2e}
\usepackage{achemso}
\setkeys{acs}{maxauthors = 0}      
\setkeys{acs}{articletitle = true} 

\graphicspath{ {./Images/} }

\begin{document}

\title{Room temperature control of axial and basal antiferromagnetic anisotropies using strain}

\maketitle

\noindent
\author{Jack Harrison,$^{1,\dag}$}
\author{Junxiong Hu,$^{2,3}$}
\author{Charles Godfrey,$^1$}
\author{Jheng-Cyuan Lin,$^1$}
\author{Tim A Butcher,$^4$}
\author{J\"org Raabe,$^5$}
\author{Simone Finizio,$^5$}
\author{Hariom Jani,$^{1,\dag,*}$}
\author{Paolo G Radaelli $^1$}
\\

\begin{affiliations}
\noindent
$^1$ Clarendon Laboratory, University of Oxford, Oxford, UK\\
$^2$ School of Physics, University of Electronic Science and Technology of China, Chengdu, China \\
$^3$ Department of Physics, National University of Singapore, Singapore\\ 
$^4$ Max Born Institute for Nonlinear Optics and Short Pulse Spectroscopy, Berlin, Germany\\
$^5$ Paul Scherrer Institut, Villigen PSI, Switzerland\\

\noindent
$^\dag$ These authors contributed equally.\\

\noindent
*Correspondence: hariom.jani@physics.ox.ac.uk;\\
 
\end{affiliations}


\medskip
\medskip 

\begin{abstract}

\noindent
\textbf{Abstract}

\noindent
Antiferromagnetic materials are promising platforms for the development of ultra-fast spintronics and magnonics due to their robust magnetism, high-frequency relativistic dynamics, low-loss transport, and the ability to support topological textures. However, achieving deterministic control over antiferromagnetic order in thin films is a major challenge, due to the formation of multi-domain states stabilised by competing magnetic and destressing interactions. Thus, the successful implementation of antiferromagnetic materials necessitates careful engineering of their anisotropy. Here, we demonstrate strain-based robust control over multiple antiferromagnetic anisotropies and nanoscale domains in the promising spintronic candidate $\alpha$-Fe$_2$O$_3$, at room temperature. By applying isotropic and anisotropic in-plane strains across a broad temperature–strain phase space, we systematically tune the interplay between magneto-crystalline and magneto-elastic interactions. We discover that strain-driven control steers the system towards an aligned antiferromagnetic state, whilst preserving topological spin textures, such as merons, antimerons and bimerons. We directly map the nanoscale antiferromagnetic order using linear dichroic scanning transmission X-ray microscopy integrated with \textit{in situ} strain and temperature control. A Landau model and micromagnetic simulations reveal how strain reshapes the magnetic energy landscape. These findings suggest that strain could serve as a versatile control mechanism to reconfigure equilibrium or dynamic antiferromagnetic states on demand in $\alpha$-Fe$_2$O$_3$, paving the way for next-generation spintronic and magnonic devices.

\end{abstract}

\vspace{5mm}
\vspace{5mm}

\newpage

Antiferromagnetic (AFM) materials are promising platforms for developing low-power spintronics and magnonics devices as they exhibit robust magnetic order and efficient ultra-fast dynamics \cite{AFM_spintronics,Wadley,AFM_spincurr,AFMSkyrm1,Hariom,Oxide_skyr_rev,SpDiff,CuMnAs}. This is due to the compensated sublattices and the corresponding absence of long-range stray fields, which makes them immune to field perturbations. AFMs host ultra-fast relativistic dynamics enabled by the exchange amplification effect, making their spin dynamics 2-3 orders of magnitude faster than ferromagnets \cite{AFM_spincurr,AFMVortex, AFMSkyrm1, AFMSkyrm2, AFM_DW_SOT,AFM_oscillator}. Moreover, insulating AFMs also host ultra-low Gilbert damping, allowing efficient long-distance spin transport with negligible Joule losses \cite{SpDiff,Fe2O3_magnons,Fe2O3_magnons_IP,YFeO3_spdiff}.

While many spintronic device concepts have been demonstrated, a major challenge is to reliably design magnetic states in AFM materials. This is because AFM thin films are subject to competing magnetic and elastic interactions, usually resulting in the emergence of multi-domain compensated AFM states that are difficult to control using conventional magnetic approaches \cite{MagElastic,SubstrClamp,Sonka_defects}. This complexity has hindered progress in key areas, such as topological spintronics and high-frequency magnonics, where uniform and tunable antiferromagnetic states are crucial. In the case of AFM topology, AFM solitons are created via the Kibble-Zurek mechanism across a symmetry-breaking phase transition \cite{Hariom,STXM,JackHolo,Fe2O3Monopoles,Francis}. This results in the random formation of solitons in a non-uniform multi-domain AFM background, making it difficult to address and use solitons individually, as required for racetrack-based applications \cite{Racetrack}. Moreover, in the case of AFM magnonics, multi-domain AFMs have markedly poor magnon transport due to spin scattering at domain boundaries \cite{Fe2O3_film_spdiff,Fe2O3_magnons_IP}. Hence, the targeted design of anisotropy in the magnetic energy landscape is an ongoing challenge, yet necessary for tuning AFM domain populations towards a mono-domain state. 

Existing strategies, such as chemical doping \cite{HDoping} and epitaxial substrate-strain \cite{FeStrain}, offer only irreversible and partial control, often constrained by substrate compatibility or limited symmetry selectivity. A scalable, reversible, and symmetry-sensitive method to tailor AFM anisotropy is therefore urgently needed. Our approach addresses this gap by leveraging strain to design antiferromagnetic anisotropy, thereby controlling the domains.

Here, we focus on the canted AFM $\alpha$-Fe$_2$O$_3$ (haematite), which is a promising quantum material with a rich AFM phase diagram \cite{MorrishBook,HDoping}, ultra-low Gilbert damping, large spin Hall magnetoresistance and domains that are switchable via spin-torques \cite{SpDiff, SpTrans, SpHall1, SpHall2, KlauiFe2O3, FeCurrent1, FeCurrent2,Oxide_skyr_rev}. $\alpha$-Fe$_2$O$_3$ thin films in the easy-plane phase have been shown to host a wide family of AFM topological textures, including (anti)merons and bimerons \cite{Hariom, STXM, Francis,Fe2O3Monopoles}, which can be reversibly nucleated at room temperature, and AFM skyrmions have been predicted to be meta-stable in the easy-axis phase \cite{MySims}. Furthermore, $\alpha$-Fe$_2$O$_3$ exhibits long-range circularly- and linearly-polarised spin-wave transport, along with ultra-fast and non-reciprocal magnons propagating up to 20 km/s \cite{SpDiff,Fe2O3_magnons,Fe2O3_film_spdiff}. Hence, $\alpha$-Fe$_2$O$_3$ is emerging as a prime candidate for magnonics and topological spintronics applications. However, $\alpha$-Fe$_2$O$_3$ typically hosts complex multi-domain AFM textures, which, in spite of the weak canted ferromagnetic moment, cannot be eliminated in remanence even after application of large magnetic fields \cite{Hariom,JackHolo,SubstrClamp}.

Strain is a versatile tool for controlling magnetic quantum materials that host spin-charge-lattice coupling, including $\alpha$-Fe$_2$O$_3$ \cite{STXM,AFMMagElast, ManipulateAFM,Oxide_skyr_rev,Strain_shape_anis,BFO_TSO,BFO_strain,MnTe_imaged}. This approach can be used to engineer the magnetic anisotropy systematically and thereby design AFM domains and topological textures. $\alpha$-Fe$_2$O$_3$ is ideally suited for investigating domain control via strain, as it hosts sizeable magneto-elastic interactions \cite{MorrishBook,SubstrClamp,Crystal_strain,Russian_strain} which could be exploited to control the local AFM order. Strain alters inter-atomic distances, and thus the strength of the magneto-crystalline interactions \cite{FeStrain,Hariom,STXM,HDoping}. It has been observed that strained epitaxial films grown on buffered corundum substrates can have a modified Morin transition temperature \cite{FeStrain}. This indicates that strain could be used to tailor AFM anisotropy. Although the studies thus far have been insightful, their approach of incorporating strain through crystallographic mismatch is fundamentally restrictive due to limited lattice parameters of compatible substrates and the formation of strain-driven defects during high-temperature growth. 

Here, we circumvent this limitation and explore large strain effects across a broad temperature-strain phase space, by exploiting free-standing AFM membranes that can accommodate sizeable strains without undergoing fracture \cite{STXM}. We employ \textit{in situ} dichroic scanning transmission X-ray microscopy (STXM) to directly map the  antiferromagnetic states. We can tune both the axial and basal anisotropies of $\alpha$-Fe$_2$O$_3$ in a controlled manner and, thereby, `design' the domain orientation and distribution. We present a Landau model which confirms that our observations are consistent with the magneto-elastic effects of strain. Micromagnetic simulations show that it is possible to design the in-plane basal anisotropy from triaxial to uniaxial by carefully tuning the relative strength of the intrinsic magneto-crystalline and strain-driven magneto-elastic interactions. Our results show that strain is a powerful and versatile technique for tailoring the nature of AFM anisotropy and the distribution of topological and trivial AFM states, as required for adaptive AFM spintronics applications.
 
\section{Results and Discussion}

\subsection{\label{sec:Fe2O3} Nature of anisotropy in $\alpha$-Fe$_2$O$_3$ }

The dominant magnetic anisotropy in $\alpha$-Fe$_2$O$_3$ is \textit{axial} due to the trigonal crystal structure (space group $R\overline{3}c$). This anisotropy emerges from the competition between magnetic-dipolar and on-site interactions favouring in-plane (IP) and out-of-plane (OOP) spin orientations, respectively \cite{MorrishBook,FeAnis,HDoping}. The temperature at which the axial anisotropy changes sign is called the Morin transition temperature ($T_M$), such that the axial anisotropy constant $K_\text{U1}>0$ for $T<T_M$ and $K_\text{U1}<0$ for $T>T_M$. This results in an easy-axis to easy-plane spin reorientation transition occurring in bulk $\alpha$-Fe$_2$O$_3$ at 260\,K \cite{MorrishBook}. Since these interactions are sensitive to temperature, pressure, and doping \cite{MorrishBook,FeAnis,HDoping}, they can be tuned easily using external perturbations and can be brought above room temperature in doped thin films and membranes \cite{Hariom, HDoping, STXM, JackHolo,Fe2O3Monopoles}. Below the Morin transition temperature ($T_\text{M}$), the N\'eel vector lies along the crystallographic $c$-axis (easy-axis state) and out-of-plane domains are separated by anti-phase domain walls. Above the transition, the N\'eel vector flips into the $ab$-plane perpendicular to the $c$-axis (easy-plane state). In this phase, in addition to the axial anisotropy, a weak triaxial \textit{basal} anisotropy, $K_\text{B}$, is present, resulting in the emergence of trigonal AFM domains separated by 120\textdegree~along with their time-reversed counterparts.

The first-order Morin transition can be understood via a simple Landau model. Here, the order parameter is two-dimensional, representing polar and azimuthal N\'{e}el vector angles. As these two parameters are perpendicular, we can expand the free energy separately in terms of the sine of the polar N\'eel vector angle ($\theta$), which varies discontinuously, as guided by the axial anisotropy, from 0 above the transition to $\pm 1$ below the transition, and the azimuthal angle $\phi$, which has preferential directions determined by the basal anisotropy.

In the absence of external strain and up to 6\textsuperscript{th} order, the free energy ($F$) can be written as \cite{FeAnis}
\begin{equation}
F = K_\text{U1}\sin^2(\theta) + K_\text{U2}\sin^4(\theta) + (K_\text{U3} + K_\text{B} \cos^2(3\phi))\sin^6(\theta) + ....
\label{eqn:LandauModel}
\end{equation}
where typically the coefficient of the 2\textsuperscript{nd} order term is taken to be temperature dependent, such that close to the transition $K_\text{U1} \approx A_0(T_\text{M} - T)$. The sign change in this lowest-order term drives the phase transition at $T = T_\text{M}$. The Morin transition is a nucleation-driven hysteretic first-order phase transition \cite{MorrishBook,Hariom,STXM}. The higher-order coefficients ($K_\text{U2}, K_\text{U3}, K_\text{B}$) are assumed to be temperature-independent close to the transition. The coefficients correspond to the anisotropy constants of the material, with $K_\text{U1}, K_\text{U2}, K_\text{U3}$ acting as second-, fourth-, and sixth-order anisotropies, respectively. $K_\text{B}$ represents the strength of the basal plane anisotropy \cite{FeAnis, MorrishBook}. The azimuthal N\'eel vector angle $\phi$ only enters the expression through the sixth-order term and is taken relative to the $a^*$ crystallographic axis \cite{Francis, FeAnis}.

\subsection{\label{sec:Biax} Effect of isotropic strain on the axial anisotropy}

\begin{figure}[!htb]
\centering
\includegraphics[width=\textwidth]{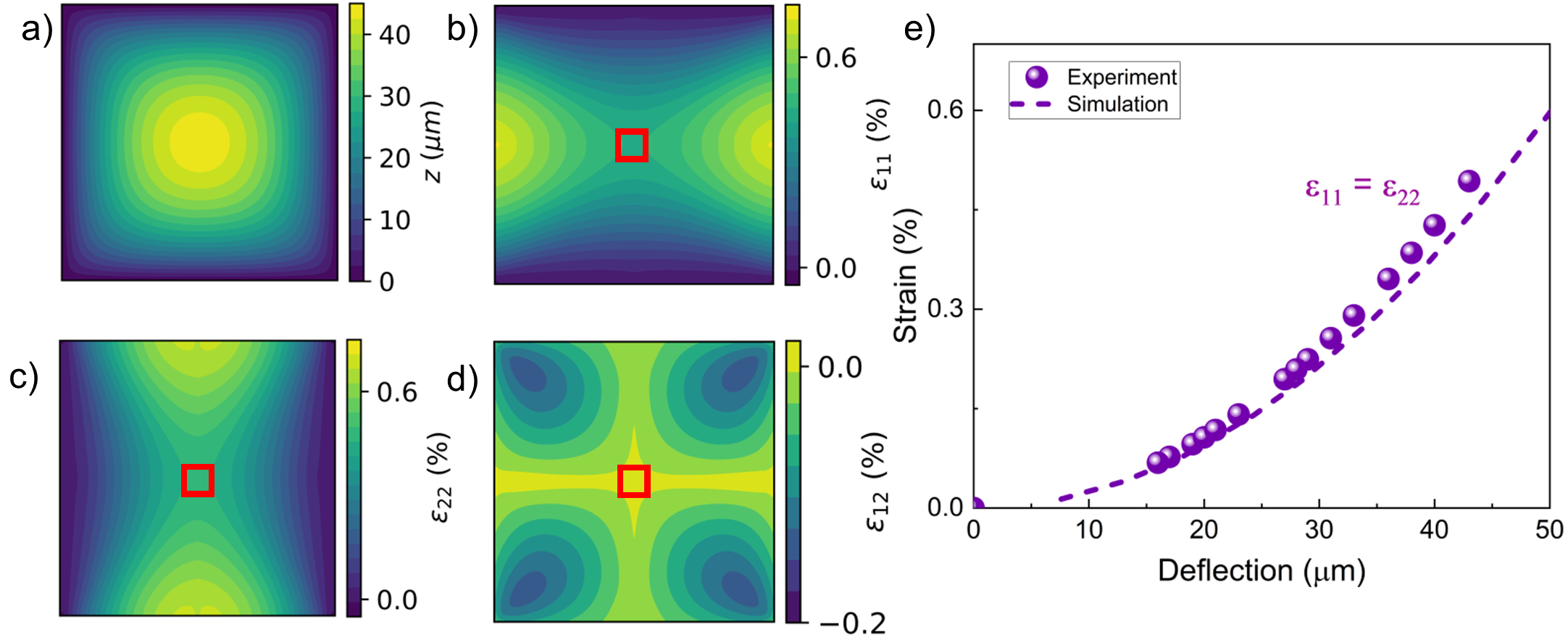}
\caption{\textbf{Strain distribution in a square membrane.} a) Example of a simulated deflection profile for a square membrane (dimensions: 1$\times$1\,mm$^2$). b-d) Corresponding maps of the strain component (b) the horizontal component $\varepsilon_{11}$, (c) the vertical component $\varepsilon_{22}$, and (d) the shear component $\varepsilon_{12}$. The red box marks the 50x50~$\mathrm{\mu}$m region about the centre within which imaging was performed. The strain is roughly uniform $\varepsilon_{11} \approx \varepsilon_{22}$ close to the centre of the membrane, and the shear component $\varepsilon_{12}$ is  negligible. e) The symmetric membrane strain at the centre of a square membrane estimated from deflection using equation \ref{eqn:StrainCali}. The dashed line is the simulated symmetric strain as a function of the membrane deflection.}
\label{fig:SqStrain}
\end{figure}

We first study the effect of strain on the axial anisotropy. To strain $\alpha$-Fe$_2$O$_3$ membranes isotropically, we employed a gas cell setup (see methods) that exploits a pressure differential to apply an approximately isotropic in-plane tensile strain to an $\alpha$-Fe$_2$O$_3$ membrane mounted on a thin square Si$_3$N$_4$ window. The strain profile was calibrated by measuring the membrane deflection as a function of pressure and comparing an analytical model of the strain with a series of finite-element simulations (see methods), as shown in figure \ref{fig:SqStrain}. We expect the strain to increase quadratically as a function of deflection from the mechanical model (equation \ref{eqn:StrainCali}, \cite{GasCell1}). The simulated in-plane strain elements $\varepsilon_{11}$ and $\varepsilon_{22}$ are symmetric near the center of the square membrane.

We collected a series of linear dichroic X-ray images (see methods) at different temperatures and gas pressures (\textit{i.e.} strain), whilst monitoring the domain populations across the Morin transition. In this imaging mode, the out-of-plane and in-plane AFM domains can be distinguished through their linear dichroic contrasts, represented in figure 2a-d as purple and yellow/orange, respectively. While the in-plane contrast changes due to the rotation of the sample azimuth or the in-plane X-ray polarization, the out-of-plane contrast remains invariant.

\begin{figure}[!htb]
\centering
\includegraphics[width=\textwidth]{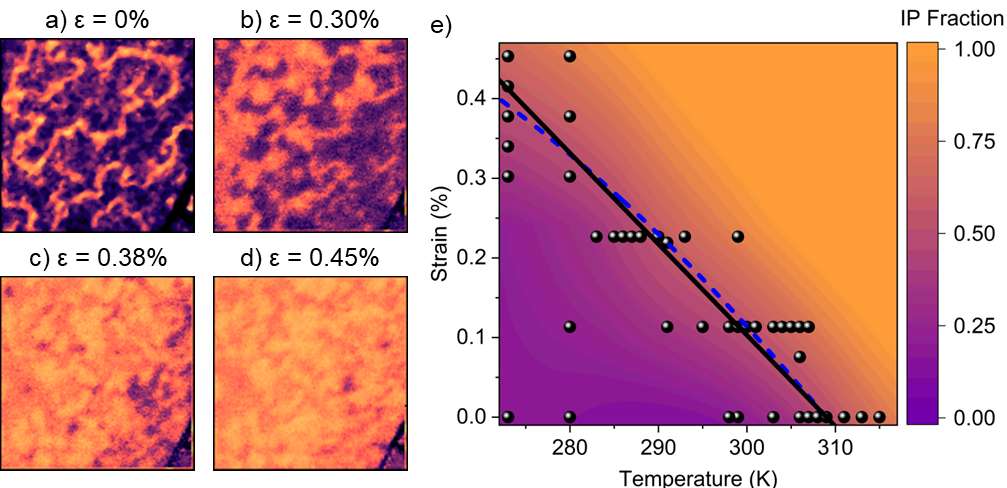}
\caption{\textbf{Effect of isotropic strain.} (a-d) 10x10\,$\mu$m XMLD-STXM images obtained at 280\,K on a membrane with an inherent (zero-strain) transition close to 309\,K. Purple and orange colours correspond to OOP and IP N\'eel vector orientations respectively. (e) Phase diagram of the magnetic state as a function of temperature and strain. The solid black line is a linear fit of the transition temperature as a function of applied strain. The dashed blue curve is a higher-order (cubic) fit that better matches the data at higher strain. The black circles indicate the positions in the phase diagram corresponding to experimental data.}
\label{fig:ST_Phase}
\end{figure}

An example set of strain-dependent images, collected below room temperature, is shown in figure \ref{fig:ST_Phase}a-d. We observe that tensile strain increases the in-plane domain fraction by first widening the anti-phase domain walls and subsequently nucleating in-plane domains. At larger strains, the sample fully transitions from the out-of-plane phase to an in-plane state. Further measurements were performed across a broad range of temperatures to obtain the full temperature-strain phase diagram, which is presented in figure \ref{fig:ST_Phase}e. Below the phase transition, the system consists of time-reversed OOP domains (purple) separated by IP domain walls (orange). Crossing the transition, the domain walls expand, increasing the IP fraction, with the transition point defined here when there is a 50\% propensity of IP and OOP domains. Above the transition, IP domains become dominant and the only OOP regions are found at the core of topological textures \cite{Francis, Hariom, STXM}, which shrink with increasing anisotropy such that they tend to zero far above the transition and the system become wholly in-plane. These results demonstrate that there is a systematic reduction in the Morin transition temperature as a function of isotropic tensile strain. In fact, the transition can be crossed completely \textit{athermally} at room temperature. This confirms a systematic strain-driven tuning of both the strength and sign of the axial antiferromagnetic anisotropy responsible for the Morin transition.

In the presence of in-plane strain, equation \ref{eqn:LandauModel} can be modified by introducing strain dependence in the axial anisotropy term. The magneto-elastic tensor and associated energy is well-known for haematite \cite{MorrishBook, MagRestConsts1, MagRestConsts2}. Neglecting any shear strain terms (shown to be very small here, figure 1d, 3d) the magneto-elastic contribution to the free energy can be written as (see Supplementary Information S5):

\begin{equation}
F_{\text{ME}} = \frac{1}{2}\sin^2(\theta) \left[K_\text{S1}(\varepsilon_{11}+\varepsilon_{22}) + K_\text{S2}(\varepsilon_{11}-\varepsilon_{22})\cos{2\phi} \right].
\label{eqn:StrainME}
\end{equation}

\noindent 
where $K_\text{S(1,2)}$ are magneto-elastic constants for haematite. From equation \ref{eqn:StrainME}, we can see that the strain acts at the second order in $\sin(\theta)$ in the Landau model of the phase transition. Specifically, equation \ref{eqn:StrainME} has two terms: an isotropic strain effect that acts symmetrically on the in-plane domains ($K_\text{S1}$) and an anisotropic term that acts as an effective induced uniaxial anisotropy ($K_\text{S2}$). For the case with square membranes, the strain is expected to be symmetric close to the membrane centre such that $\varepsilon_{11}\approx\varepsilon_{22}$ and we can neglect the asymmetric term (as confirmed in figure 1), although this second term will become important later. 

Taking only the term symmetric in strain we can modify the Landau model (equation \ref{eqn:LandauModel}) such that $K_\text{U1}=A_0(T_\text{M}-T)+K_\text{S1} \varepsilon$, where $\varepsilon = \frac{1}{2}(\varepsilon_{11}+\varepsilon_{22})$ is the symmetric strain. This evolution is fully consistent with our experimental results, where tensile strain is found to systematically reduce the $T_\text{M}$, indicating that $K_\text{S1}<0$. The solid black curve in figure \ref{fig:ST_Phase} is a linear fit to the strain-dependent transition temperature based on equation \ref{eqn:StrainME} for the case of isotropic strain, demonstrating that this function explains the data adequately. The dashed blue curve shows a higher-order cubic fit, which is necessary to model the effect at larger values of strain. In general, higher-order strain terms will be present and become more relevant at larger strain values. The inclusion of even-order (quadratic, quartic) terms has a negligible effect on the fit, reflecting the inverse effect expected from compressive compared to tensile strains \cite{FeStrain}. A higher-odd-order fit was not sufficiently constrained by the data, and is thus not discussed further.

\subsection{\label{sec:Uniax} Effect of anisotropic strain on the basal-plane anisotropy}

As previously mentioned, in addition to the dominant axial anisotropy, a weak basal anisotropy is also present in $\alpha$-Fe$_2$O$_3$ \cite{MorrishBook}, corresponding to the $K_B$ term in equation \ref{eqn:LandauModel}. This term is responsible for the formation of distinct trigonal in-plane AFM domains and their time-reversed counterparts in the easy-plane state above the Morin transition \cite{Hariom,Francis}. At the microscopic level, the origin of this basal anisotropy is related to the 3-fold rotational symmetry of the Fe-cations encased in face-sharing O-octahedra \cite{FeAnis}. Hence, one can hypothesise that anisotropic in-plane distortions would lift the sixfold degeneracy enforced by the basal anisotropy via the second term of equation \ref{eqn:StrainME}.

\begin{figure}[!t]
\includegraphics[width=\textwidth]{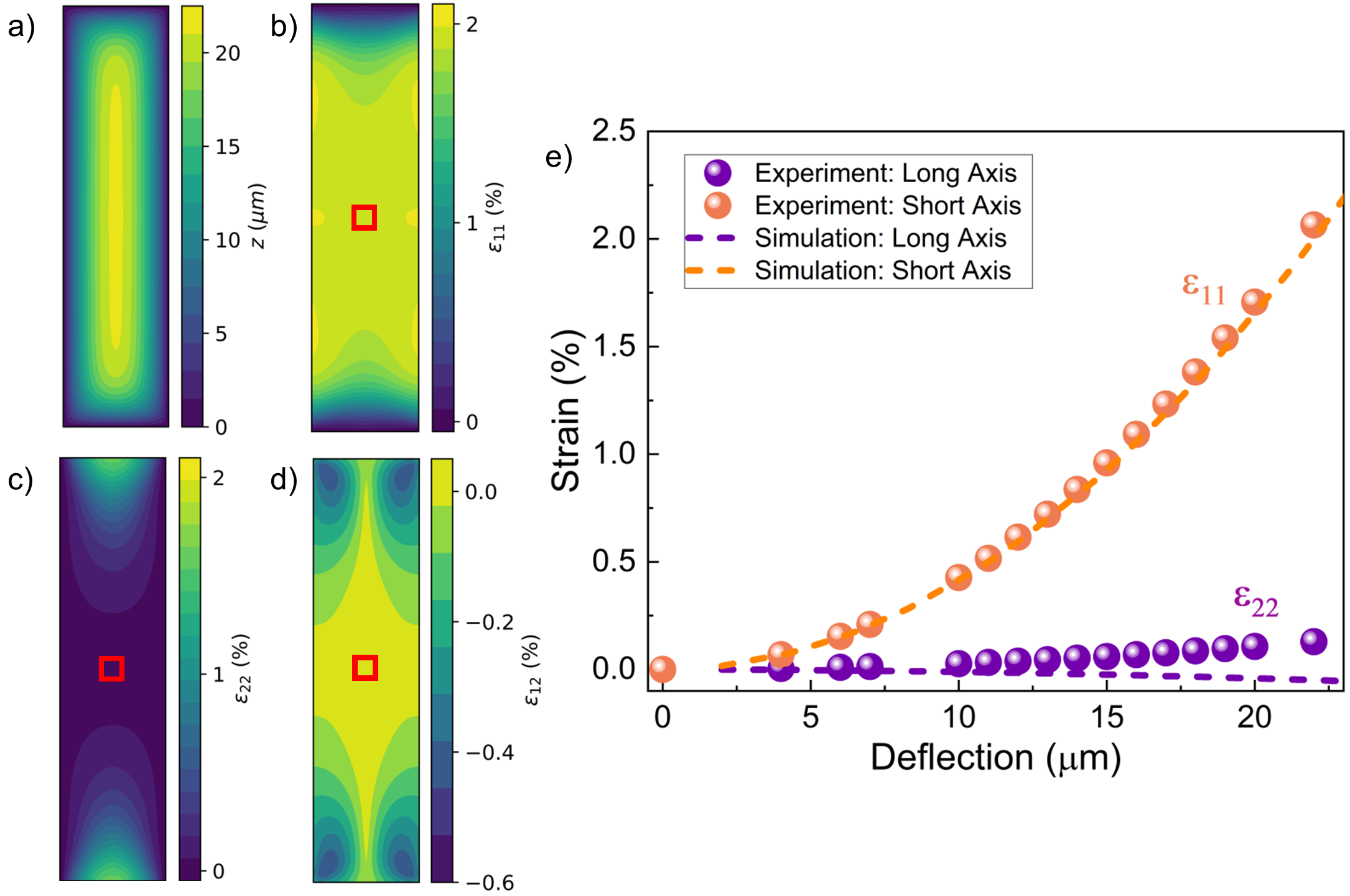}
\caption{\textbf{Strain distribution in a rectangular membrane.} a) Example of a simulated deflection profile for a rectangular membrane (dimensions: 1$\times$0.25\,mm$^2$). b-d) Corresponding maps of the strain component across (b) the short, high-strain axis $\varepsilon_{11}$, (c) the long, low-strain axis $\varepsilon_{22}$, and (d) the shear component $\varepsilon_{12}$. The red box marks the 50x50~$\mathrm{\mu}$m region about the centre within which imaging was performed. The strain is roughly spatially uniform (but anisotropic) close to the centre of the membrane. e) Strain components at the membrane centre estimated from the measured deflection using equation \ref{eqn:StrainCali}. The dashed lines are the simulated strain components as a function of the membrane deflection.}
\label{fig:RectStrain}
\end{figure}

In order to demonstrate the effects of anisotropic (uniaxial) strain on the AFM order, we investigated an $\alpha$-Fe$_2$O$_3$ membrane on a rectangular Si$_3$N$_4$ holder (see methods). The deflection-strain calibration from both finite-element simulations and measurements is presented in figure \ref{fig:RectStrain}. Pressurising a rectangular membrane results in highly anisotropic tensile strains, such that the strain along the shorter axis is much larger than its counterpart along the longer axis, \textit{i.e.} $| \varepsilon_{11}/\varepsilon_{22}|>>1$. Both the $\varepsilon_{11}$ and $\varepsilon_{22}$ strain components obey the same quadratic scaling with membrane deflection (equation \ref{eqn:StrainCali}, methods), but have drastically different strengths due to the different radius of curvature along the two axes. In fact, we find from the finite-element simulations that $\varepsilon_{22}$ is expected to be both small and negative (compressive), contrary to $\varepsilon_{11}$ that is large and positive (tensile), at the membrane centre (figure \ref{fig:RectStrain}e).

While tensile strains in this experimental geometry also suppress the axial anisotropy, as observed in the previous section (see Supplementary Information, figure S4), new strain effects manifest due to changes in the basal anisotropy. Hence, in the following discussion we focus on the effect of the anisotropic strain on trigonal domains \emph{above} the Morin transition. To identify strain-driven in-plane domain reorganization, we constructed AFM vector maps as follows: the incident X-ray linear polarization was rotated in steps of 15\textdegree~ and a total of 7 images were collected at polarization angles between 0\textdegree~ and 90\textdegree~ to the horizontal. These were then used to generate a vector map at three different pressures, following the method we developed for transmission-based X-ray imaging (see methods) \cite{STXM}.

\begin{figure}[!htb]
\centering
\includegraphics[width=\textwidth]{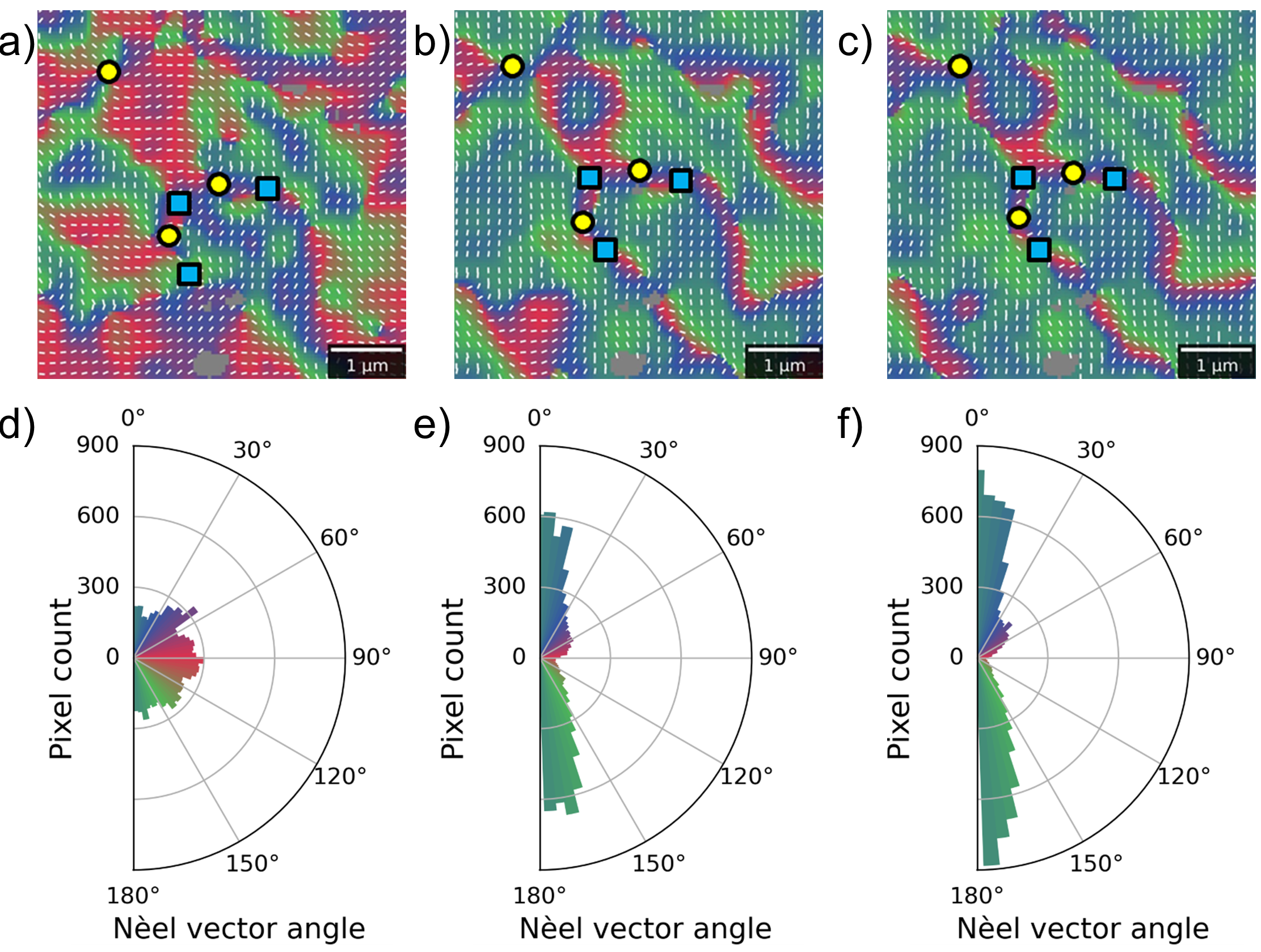}
\caption{\textbf{Effect of anisotropic strain.} (a-c) Vector maps and (d-f) corresponding pixel-wise distribution of phase angle in the same region of an $\alpha$-Fe$_2$O$_3$ membrane near the centre of a rectangular Si$_3$N$_4$ holder. The images correspond to three different applied pressures: (a,d) 0\,mbar unstrained state, (b,e) 400\,mbar, $\varepsilon_{11} = 0.35\%$ and (c,f) 800\,mbar, $\varepsilon_{11} = 1.23\%$. Red/green/blue colours in both the vector maps and pole plots correspond to the three trigonal domains and the local N\'eel vector orientation is indicated by white bars in the vector maps. Regions identified as topological merons and antimerons are shown by yellow circles and blue squares, respectively. The grey regions are defects on the sample surface. The white scale bar is 1\,$\mathrm{\mu m}$ long.}
\label{fig:UniStrain}
\end{figure}

As shown in figure \ref{fig:UniStrain}, applying anisotropic strain to the sample has a drastic effect on the domain populations. This is emphasised even more clearly by the corresponding pole plots, which are angular histograms of the AFM phase angle across each image using the same colour scheme as the vector map images. In the unstrained (0\,mbar) state, there is a roughly equal proportion of the three trigonal domains (colours: red, green and blue). The intermediate regions (intermixed colours: orange, purple, and turquoise) are more than 15\textdegree~ from any of the principal trigonal axes and make up a proportion of the overall structure, effectively acting as wide in-plane domain walls separating the trigonal domains.
This also demonstrates that the basal plane anisotropy is indeed rather weak \cite{FeAnis, MorrishBook, Hariom}. Under application of 400\,mbar and then 800\,mbar gas pressure, with the largest component of tensile strain along the horizontal axes of these figures, the `red' domains shrink drastically, whilst the orthogonally oriented domains grow. Specifically, the intermediate (turquoise) regions perpendicular to the high-strain axis are significantly enhanced, signifying an induced uniaxial anisotropy perpendicular to the strain competing with the basal anisotropy. 

To gain a quantitative understanding of this phenomenon, we consider the effect of anisotropic strain on the domain distribution through the magneto-elastic interactions \cite{SubstrClamp,Crystal_strain,Russian_strain} as described in equation \ref{eqn:StrainME}. The 2\textsuperscript{nd} order term in the free energy therefore becomes:
\begin{equation}
\left(A_0(T_\text{M} - T) + \frac{1}{2}\left[K_\text{S1}(\varepsilon_{11}+\varepsilon_{22}) + K_\text{S2}(\varepsilon_{11}-\varepsilon_{22})\cos{(2\phi-\chi)}\right]\right)\sin^2(\theta).
\label{eqn:Triaxial}
\end{equation}

 This equation shows that in addition to the strain-induced reduction of T$_\text{M}$ found in the case of isotropic strain above, the anisotropic strain term  (when $\varepsilon_{11} \neq \varepsilon_{22}$) induces an effective uniaxial anisotropy within the basal plane, along a special direction relative to the strain axis. Here we have introduced a small angular offset $\chi$ between the crystallographic axis and the experimental x-axis, which also corresponds to the high-strain axis, due to a small misalignment of the membrane during transfer. The asymmetric term $\cos(2\phi-\chi)$ ensures that the strain-induced in-plane anisotropy reverses under a 90\textdegree~rotation of the strain axis \cite{Russian_strain}. 
 
 Whether the favoured domains are parallel or perpendicular to the high-strain axis ($\varepsilon_{11}$) depends on the sign of $K_\text{S2}$. We observe in figure \ref{fig:UniStrain} that the red domains parallel to the high-strain axis are disfavoured at the cost of the turquoise domains, suggesting that the N\'eel vector hard-axis is along the high tensile strain and that the sign of $K_\text{S2}$ is negative. This is consistent with the magneto-elastic constants reported for $\alpha$-Fe$_2$O$_3$, wherein the authors also find that $K_\text{S2}$ is negative \cite{MagRestConsts1, MagRestConsts2}.

In the ideal scenario, one would expect that increasing the strain-induced uniaxial anisotropy would cause the favoured domains to expand, whilst the disfavoured ones are removed entirely, if the anisotropy is strong enough. This is the case in many regions of figure \ref{fig:UniStrain}, where the turquoise domains at 90\textdegree~ to the high-strain axis grow at the cost of the red ones parallel to the strain. Furthermore, as the strain increases, the favoured domains tilt towards the uniaxial anisotropy axis and should fully collapse towards it under extreme strain. 

Interestingly, notable exceptions to this model are the regions around topological cores, where the disfavoured domains are necessarily pinned by the topological winding of the textures. This makes such topological objects particularly useful for studying domain distributions as a function of anisotropy, since they act as points of intersection between all of the trigonal domains and their time-reversed counterparts \cite{Francis}. In figure \ref{fig:UniStrain}a, merons and antimerons are identified as regions where there is a 360\textdegree~ winding of the N\'eel vector about a point.Under the application of strain, the topological textures are found to be robust, which is consistent with the winding of the domains around the (anti)meron core being topologically protected. It is noteworthy that these topological textures appear to be linked by tightly bound `strings'. In fact, compared to the turquoise sectors, the area of the red sectors is significantly reduced, so that they in effect become the `strings' connecting the topological cores -- see figure \ref{fig:UniStrain}b,c. In order for topological bimerons, which are tightly bound meron-antimeron pairs, to be useful for racetrack-based device applications, they are required to exist in a largely uniform background. These results indicate that straining AFM layers could be a promising pathway to discovering such isolated textures.

\subsection{\label{sec:MicroSims} Micromagnetic simulations}

\begin{figure}[!htb]
\centering
\includegraphics[width=\textwidth]{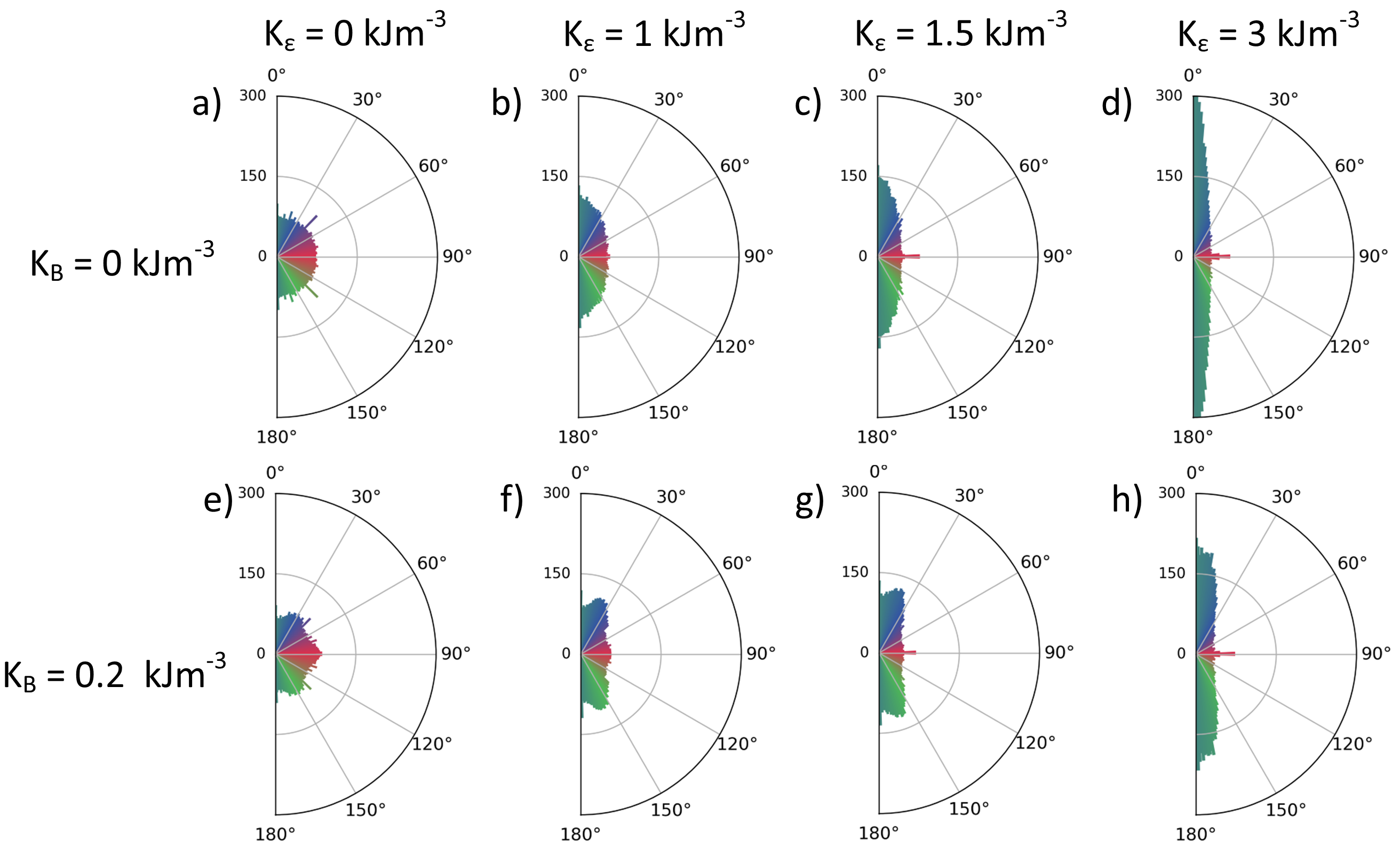}
\caption{\textbf{Micromagnetic simulations.} Pole plots of the angular distribution for simulated merons with different values of strain-induced anisotropy $K_{\varepsilon}$ and basal plane anisotropy $K_\text{B}$.}
\label{fig:ExpSim}
\end{figure}

To elucidate the effects of strain on the in-plane domain distribution, we conducted a series of micromagnetic simulations using the MuMax3 software \cite{mumax1, mumax2, mumax3}, following the model of A-type AFMs presented in Ref \cite{MySims}. For this study, it was important to appropriately initialise the simulation to obtain quantitative results. We found that a generic collection of in-plane domains evolved irregularly under strain whilst being sensitive to initial conditions, as meta-stable domains were pushed out of the finite simulation box. Therefore, we chose to initialise the simulation with all in-plane sectors populated equally and fully wound around a common topological (anti)meron core. As discussed above, topological (anti)merons always stabilise a full set of AFM domains winding around the core, so there is no possibility for a domain to drift entirely out of the simulation box. Hence, relative populations of majority/minority domains can be investigated under different anisotropies. Moreover, (anti)merons are experimentally relevant because they are found to be ubiquitous in $\alpha$-Fe$_2$O$_3$ thin film and membrane samples \cite{Hariom,STXM, JackHolo}. 

The relevant parameters in our simulation were the AFM exchange interaction ($A$), the axial anisotropy along the c-axis responsible for the Morin transition ($K_\text{U}$), a basal plane anisotropy ($K_\text{B}$) and an additional uniaxial anisotropy ($K_{\varepsilon}$) to model the effects of strain following equation \ref{eqn:Triaxial} (for full details of the micromagnetic implementation see Supplementary Information S6). We simulated topological merons to compare their winding directly with those found experimentally in figure \ref{fig:UniStrain}, with the simulation images shown in Supplementary Information figure S6.

As shown in figure \ref{fig:ExpSim}, in the absence of applied strain the simulated basal plane anisotropy ($K_\text{B}>0$) enforces a weakly trigonal structure, distorting the fully easy-plane state and preferentially stabilising N\'eel vector orientations along the three in-plane axes (red-green-blue). This corresponds well to the real winding around similar experimentally observed textures in the unstrained state, as shown in figure \ref{fig:UniStrain}a and Ref \cite{Hariom}. Next, upon introducing the strain-induced uniaxial anisotropy in the absence of basal anisotropy, figure \ref{fig:ExpSim}b-d, we observe the creation of an easy axis orthogonal to the horizontal direction of the strain. Finally, in the presence of both strain-induced uniaxial anisotropy and basal anisotropy, we observe a competition between the two interactions, such that two basal axes are slightly biased over the third (figure 5f,g). At higher values of strain, the system collapses towards a uniaxial state (figure 5h). These results are fully consistent with the effect of the strain-induced anisotropy observed experimentally, figure \ref{fig:UniStrain}.

Both of these interactions alter the population of in-plane domains and, in doing so, also distort the local winding around topological cores (see figure S6), whilst preserving the topological winding number \cite{Francis,Hariom}. Thus, we have demonstrated that introducing the basal anisotropy into our micromagnetic simulations recreates the trigonal domain patterns seen in the absence of strain. Furthermore, introducing an additional uniaxial anisotropy energy term models the strain-induced reconfiguration of the trigonal domains. This simulation scheme therefore represents both an advancement over previous micromagnetic simulations in this material class \cite{MySims} and a demonstration that a strain-induced uniaxial anisotropy is an excellent explanation for the observed effects in figure \ref{fig:UniStrain}.

\section{Conclusion and Outlook}
In summary, we demonstrate robust, room temperature strain control of both axial and basal antiferromagnetic anisotropies in thin layers of the model antiferromagnet $\alpha$-Fe$_2$O$_3$. The use of free-standing crystalline membranes allowed us to design the magneto-elastic interactions across a broad temperature–strain phase diagram. Isotropic tensile strain was shown to change the sign of the axial anisotropy and, thereby, favour the topologically-rich easy-plane state compared to the easy-axis state. In contrast, in-plane anisotropic strain was shown to break the symmetry of the basal plane anisotropy, from effectively triaxial to uniaxial. This was evidenced through a strain-induced reorientation of the domain populations toward a nearly aligned antiferromagnetic state. A phenomenological Landau model, extended to include magneto-elastic contributions, and micromagnetic simulations revealed how strain reshapes the magnetic free energy landscape, providing a framework to explain the observed anisotropy reconfigurations. Crucially, topological antiferromagnetic textures present in the unstrained state are shown to be robust even in the presence of large strains, albeit with reconfiguration of the AFM domains surrounding the cores. 

Our findings demonstrate that strain is a versatile and robust tuning parameter for deterministic engineering of multiple anisotropies and domain orientations in $\alpha$-Fe$_2$O$_3$. Such strains could be applied statically or dynamically using carefully selected crystalline or piezoelectric substrate platforms to make practical devices, further confirming that $\alpha$-Fe$_2$O$_3$ is an excellent platform for spintronics applications. Specifically, strained $\alpha$-Fe$_2$O$_3$ layers could be relevant for topological devices, such as race-tracks and reservoirs that require topological textures in a controlled background, either individually or coexisting with other textures \cite{AFM_SkyrmMotion,BimChaos,AFM_reservoir}, as well as for high-frequency magnonic devices, including spin-wave gates, oscillators and rectifiers \cite{AFM_oscillator,AFM_rectifier,AFM_droplets}.

\section{\label{sec:Methods} Methods}

\noindent
\textbf{Sample Fabrication:} $\alpha$-Fe$_2$O$_3$ membranes were grown using pulsed laser deposition on SrTiO$_3$ (111) substrates with a water-soluble Sr$_3$AL$_2$O$_6$ sacrificial layer and an intermediate SrTiO$_3|$LaAlO$_3$ ``buffer'' layer. They were then detached via an indirect water etching and liftoff transfer process, for full details see \cite{STXM,JackHolo}. Post water etching, membranes were transferred onto 50\,nm thick Si$_3$N$_4$ held in 5x5\,mm Si frames. Si$_3$N$_4$ windows were either square (1$\times$1\,mm$^2$) or rectangular (1$\times$0.25\,mm$^2$), allowing for either isotropic or uniaxial strain to be applied, respectively.

\vspace{5mm}
\noindent
\textbf{\textit{In situ} X-ray imaging:} Scanning transmission X-ray microscopy (STXM) with X-ray magnetic linear dichroism (XMLD) was used to study the antiferromagnetic structure in these membranes, following the methodology developed in our previous paper \cite{STXM}. The chamber was fitted with a custom gas cell (see figure S1) \cite{GasCell1,GasCell2,FlexExample}, where the transferred $\alpha$-Fe$_2$O$_3$ membrane integrated on a Si$_3$N$_4$ holder was used as one of the two sealing membranes and the other was a blank membrane. The inside of the cell was pressurised by a He gas input with flow rates 0-100\,cc/min. A PID controlled needle valve connected to a vacuum pump was used to set and maintain the pressure in the range 0-800\,mbar with a stability better than $\pm$1\,mbar, all controlled and monitored by a computer interface. The endstation outside the cell was held at a vacuum better than $10^{-5}$\,mbar and this pressure difference causes the membrane and sample to flex. The membrane deflection was measured by the change in the focal distance of the Fresnel zone plate, with accuracy $\pm$2\,$\mathrm{\mu m}$. To measure this, we refocused the microscope at each pressure and recorded the focal distance. The change in focal distance compared to the unstrained state then gives the membrane deflection. The induced flexure was found to be temperature-independent for a given pressure within the ranges used here.

Using the measured deflection as a function of pressure, the strain can be estimated close to the centre of the membrane following the mechanical model presented in Ref \cite{GasCell1}. For a circular membrane of radius $r$ undergoing a deflection of height $h$, the radial component of strain is
\begin{equation}
\varepsilon_r = \frac{2h^2}{3r^2}.
\label{eqn:StrainCali}
\end{equation}
The above calibration is expected to be valid only close to the centre of the membrane where the deflection is approximately circularly symmetric, hence all measurements were conducted within 50 $\mu$m of the membrane centre (red squares in figures 1 and 3). The full calibration of the membrane strain as a function of the deflection for an example square holder is shown in figure \ref{fig:SqStrain}\,e and for a rectangular holder in figure \ref{fig:RectStrain}\,e.

\vspace{5mm}
\noindent
\textbf{Strain modelling:} In order to confirm the calibration of the strain components in the $\alpha$-Fe$_2$O$_3$ membranes as a function of deflection, a finite element model of the sample was created. The sample was modelled as a membrane with no bending stiffness and fixed boundary conditions on all four edges \cite{MechSims}. It was constructed of three layers rigidly adhered to each other: $\alpha$-Fe$_2$O$_3$, LaAlO$_3$ and Si$_3$N$_4$ with thicknesses 30 nm, 10 nm and 50 nm respectively and with mechanical properties obtained from the literature \cite{LAO_YM, SiN_YM, MorrishBook}. The full elastic tensor was used to model the $\alpha$-Fe$_2$O$_3$ and LaAlO$_3$ layers due to their known anisotropic stress-strain relations while the Si$_3$N$_4$ layer was modelled as behaving as an isotropic solid due to its polycrystalline nature. The Si$_3$N$_4$ was modelled with an in-plane tensile pre-stress (arising from growth \cite{SiN_Growth}) of 350 MPa. A stationary solution to the membrane deformation for the rectangular and square geometries at different applied hydrostatic pressures was found.

To address whether the strain applied to the Si$_3$N$_4$ membrane is fully transferred to the $\alpha$-Fe$_2$O$_3$ layer, we calculate an order-of-magnitude estimate of the critical slipping strain using the formula $\varepsilon = cL/2Et$ \cite{SlidingGraphene, 2DAdhesion}. Here, $\varepsilon$ is the critical strain, $c$ is the critical van der Waals shear stress at the interface, $L$ is the length of the hematite flake, $E$ is the Young’s modulus and $t$ is the thickness. Taking the values $E$=360×10$^{9}$ \cite{Fe2O3_YM}, $c$= 1.64×10$^6$ \cite{InterlayerStress}, $L$=1\,mm, $t$=40\,nm gives a critical strain of 5.7\%. The maximum experimental strain values in our study are well below this threshold. Although $c$ is taken from an interface with a van der Waals material, this analysis provides a reasonable order-of-magnitude estimate that the applied strains remain within the adhesion limit. We note that this estimate represents a lower bound, since typical membrane dimensions ($L$ approx. 5 mm) would further increase the critical strain, comfortably accounting for parameter variations.

\vspace{5mm}
\noindent
\textbf{Micromagnetic simulations:} Following the approach presented in Ref \cite{MySims}, we use a sublattice magnetisation $M_S = $ 920\, kAm$^{-2}$, an exchange parameter $A= 17$\,pJm$^{-1}$, and uniaxial anisotropy just above the Morin transition $K_\text{U}$ = 15\,kJm$^{-3}$. These are consistent with previous simulations in this material and the fundamental exchange and anisotropy parameters \cite{Hariom, Jacopo, MorrishBook, FeAnis, BPAnis, CoeyBook, HematiteJvals}. As the real basal plane anisotropy is incredibly small (on the order of 1\,Jm$^{-3}$ ),\cite{BPAnis, FeAnis} this is not sufficient to induce a triaxial anisotropy in the simulations with the energy minimisation criteria. To model the strain effects in our samples, we consider two scenarios - (i) easy plane anisotropy (0 kJm$^{-3}$), figure 5a-d, and (ii) relatively stronger triaxial anisotropy (0.2 kJm$^{-3}$), figure 5e-h. In both cases, the resulting antiferromagnetic domain configurations in the highly strained state are qualitatively consistent with the experimental observations shown in figure 4f. This agreement supports the applicability of our micromagnetic framework to systems possessing intermediate values of $K_\text{B}$, as is expected for the samples studied here experimentally.

\medskip
\noindent
\section{Acknowledgements} 
This research was supported by the Engineering and Physical Sciences Research Council grant (EP/M020517/1) and the Oxford-ShanghaiTech collaboration project. H.J. acknowledges the support of the Royal Society URF Grant (URF/R1/241120) and the MSCA Fellowship under the UKRI Horizon Europe Guarantee Funding (EP/X024938/1). J.H. was supported by the EPSRC (DTP Grant No. 2285094). T.A.B. acknowledges funding from the Swiss Nanoscience Institute (SNI). Part of this work was performed at the PolLux (X07DA) and the SIM (X11MA) beamlines of the Swiss Light Source, Paul Scherrer Institut, Villigen, Switzerland. The PolLux endstation was financed by the German Bundesministerium für Bildung und Forschung (BMBF) under contracts 05K16WED and 05K19WE2. We thank A. Ariando (NUS) and the Singapore National Research Foundation project (NRF-CRP15-2015-01) for support with materials. We thank Z.S. Lim (NUS) for test materials. We thank A. Hrabec (PSI) for sample preparation in the SQUID.

\section{Ethics declarations}
The authors declare no competing interests.

\section{Supplementary Information}
Supplementary information provides further experimental data, theoretical analysis, and simulation results.

\section{Data availability}
The data and analysis code that support the findings of this study are available from the authors upon reasonable request.

\newpage
\medskip

\bibliographystyle{chem-acs, articletitle=true}
\bibliography{reflist}

\end{document}